\documentclass[11pt,letterpaper]{article}
\usepackage{epsfig,rotating,setspace,latexsym,amsmath,epsf,amssymb,bm}
\usepackage{cite,graphicx,authblk,color,subfigure}

\title{The Capacity of Cache Aided Private Information Retrieval 
\author{Ravi Tandon}
\affil{Department of Electrical and Computer Engineering\\
University of Arizona, Tucson, AZ, USA.\\
E-mail: tandonr@email.arizona.edu}}

\newtheorem{Theo}{Theorem}

\newtheorem{Lem}{Lemma}
\newtheorem{Cor}{Corollary}

\begin{document}
\maketitle
\newcommand\blfootnote[1]{%
  \begingroup
  \renewcommand\thefootnote{}\footnote{#1}%
  \addtocounter{footnote}{-1}%
  \endgroup
}

\blfootnote{This work was supported in part by the NSF Grant  CAREER-1651492.}

\thispagestyle{empty}
\vspace{-1.5cm}
\begin{abstract}
The problem of cache enabled private information retrieval  (PIR) is considered in which a user wishes to privately retrieve one out of $K$ messages, each of size $L$ bits from $N$ distributed databases. The user has a local cache of storage $SL$ bits which can be used to store any function of the $K$ messages. The main contribution of this work is the exact characterization of the capacity of cache enabled PIR as a function of the storage parameter $S$. In particular, for a given cache storage parameter $S$, the information-theoretically optimal download cost $D^{*}(S)/L$ (or the inverse of capacity) is shown to be equal to $(1- \frac{S}{K})\left(1+ \frac{1}{N}+ \ldots + \frac{1}{N^{K-1}}\right)$.  Special cases of this result correspond to the settings when $S=0$, for which the optimal download cost was shown by Sun and Jafar to be $\left(1+ \frac{1}{N}+ \ldots + \frac{1}{N^{K-1}}\right)$, and the case when $S=K$, i.e., cache size is large enough to store all messages locally, for which the optimal download cost is $0$. The intermediate points $S\in (0, K)$ can be readily achieved through a simple \textit{memory-sharing} based PIR scheme. The key technical contribution of this work is the  converse, i.e., a lower bound on the download cost as a function of storage $S$ which shows that memory sharing is information-theoretically optimal. 
\end{abstract}
\section{Introduction}

Private information retrieval (PIR) refers to the problem where a user wants to efficiently retrieve one message out of $K$ messages from $N$ databases with each database storing all of the $K$ messages (i.e. all of the $N$ databases are duplicates of each other) without revealing anything about the identity of the desired message. Successful PIR must satisfy two properties: first, each of the $N$ message queries respectively sent from the user to each of the $N$ databases must reveal nothing about the identity of the message being requested; and second, the user must be able to correctly decode the message of interest from the answers received from the $N$ databases. The sum of all responses from the databases to the user are considered to constitute the download communication cost. While a trivial solution is to retrieve all the messages completely, this is clearly inefficient, and the goal of PIR is to achieve the goal in an efficient manner. Since the introduction of PIR in \cite{ChorAndCompany}, this problem has received significant attention in the computer science community \cite{TrinabhGupta, Demmler, CachinAndCompany, Yekhanin}.


In a recent interesting work \cite{SunAndJaffar1}, Sun and Jafar characterized the exact information-theoretic capacity (or the inverse of download cost) of the $(N,K)$ PIR problem as $(1 + 1/N + \ldots 1/N^{K-1})^{-1}$, improving upon the previous best known achievable rate for the PIR problem \cite{NiharPIR}. 

Since the appearance of \cite{SunAndJaffar1}, significant progress has been made on a variety of variations of the basic PIR problem. We briefly describe some of these advances next. The case of $T$-colluding PIR (or TPIR in short) was investigated in \cite{SunAndJaffar2}, where any $T$ databases out of $N$ are able to collude, i.e., they can share the queries. Robust PIR, in which  any $N$ out of $M$ databases (with $N\leq M$) fail to respond was also investigated in \cite{SunAndJaffar2}, for which the capacity is found to be the same as that of TPIR. In a recent work, \cite{BanawanAndUlukus2} characterized the  capacity of PIR with byzantine databases (or BPIR), i.e., a scenario in which any $L$ out of $N$ databases are adversarial (i.e. they can respond with incorrect bits after receiving the query). The above previous works assumed the presence of replicated databases, i.e., each database stores all the $K$ messages. The capacity of PIR with databases storing MDS coded messages was considered in \cite{Salim} and the capacity was subsequently characterized by Banawan and Ulukus in \cite{BanawanAndUlukus}. This setting was further investigated for the scenario where any $T$ out of $N$ databases can collude, an aspect termed MDS-TPIR \cite{FreijGnilkeHollantiKarpuk, SunAndJafar3} although its capacity remains open for general set of parameters. The problem of symmetric  PIR (SPIR) was studied in \cite{SunAndJafar4}. In this setting, privacy is enforced in both directions: i.e.,  user must be able to retrieve the message of interest  privately while at the same time the databases must avoid any information leakage to the user about the remaining messages. The exact capacities for this symmetric PIR problem both  for non-coded (SPIR) and MDS-coded (MDS-SPIR) messages were characterized in \cite{WangAndSkoglund, SunAndJafar4}. The case of multi-message PIR (MPIR) was investigated in \cite{BanawanAndUlukus3, ZhangAndGennian}, in which the user wants to privately retrieve $P\geq 1$ out of $K$ messages.

 \textbf{Contribution of this work--} The focus of this work is on \textit{cache aided PIR}, a setting in which the user has a local cache of size $SL$ bits, where $L$ is the size of each message. Since there are $K$ messages, the storage parameter $S$ can take values in the range $0\leq S\leq K$. The main contribution of this work is the characterization of the capacity of cache aided PIR as a function of the storage parameter $S$. It is shown that the capacity, $C^{*}(S)$ is given by $\left((1- \frac{S}{K})\left(1+ \frac{1}{N}+ \ldots + \frac{1}{N^{K-1}}\right)\right)^{-1}$. The achievability for an arbitrary storage parameter $S$ follows by a simple \textit{memory-sharing} argument (an equivalent of the idea of time-sharing used in obtaining rate-regions for various multi-user networks) that utilize the optimal PIR schemes for extreme values of $S$, i.e., $S=0$ \cite{SunAndJaffar1}, and $S=K$. The novel aspect is the converse  which provides a lower bound on the download cost (equivalently, an upper bound on capacity), and shows that the simple memory sharing scheme is in fact optimal.

\section{Cached Aided PIR: Problem Statement}
We consider $K$ independent messages $W_1, W_2, \ldots, W_K$, each of size $L$ bits, i.e.,
\begin{align}
&H(W_1, \ldots, W_K) = H(W_1) + \ldots + H(W_K), \\
&H(W_1)=\ldots = H(W_K)= L.
\end{align}
There are $N$ distributed databases and every database stores all the messages $(W_1, \ldots, W_K)$. In PIR, the user private generates an index $\theta \in [K]$ and wishes to retrieve $W_\theta$, which keeping the index $\theta$ secret from all the databases. 

In \textit{cache enabled PIR}, the user is equipped with a local cache, denoted by a random variable $Z$. The cached content $Z$ can be any arbitrary function of the $K$ messages $(W_1, \ldots, W_K)$ and is of size $SL$ bits, i.e., 
\begin{align}
&H(Z|W_1, \ldots, W_K)= 0,\\
&H(Z)= SL,
\end{align}
where the storage parameter $S$ takes values in the range $S\in [0, K]$, with $S=0$ corresponding to no storage, and $S=K$ corresponding to the case of full storage (user can store all the $K$ messages). It is assumed that the databases know $Z$, i.e., which contents are cached by the user, and the cache contents remain fixed over time. 

Depending on the message $\theta=k$ being requested, the user generates $N$ queries $Q_1^{[k]}, \ldots, Q_N^{[k]}$. The user sends the query $Q_n^{[k]}$ to the $n$th database and the $n$th database responds by an answer $A_n^{[k]}$, which is a function of $Q_n^{[k]}$, and the data stored (i.e., $W_1, W_2,\ldots, W_K)$. Hence, we have
\begin{align}\label{answer}
H(A_{n}^{[k]}|Q_{n}^{[k]}, W_1, W_2, \ldots, W_K)=0.
\end{align}
Upon receiving the answers from all the $N$ databases, and using its local cached contents $Z$, the user must be able to reconstruct the desired message $W_k$ with probability of error $P_e$. The probability of error must approach zero as $L\rightarrow \infty$, i.e., as the message size approaches infinity. In other words, through a simple application of Fano's inequality, we must satisfy
\begin{align}\label{decodability}
H\left(W_k | Z, A_1^{[k]}, \ldots, A_N^{[k]}, Q_1^{[k]}, \ldots, Q_N^{[k]}\right) = o(L),
\end{align}
where $o(L)$ represents a function such that $o(L)/L$ approaches zero as $L$ approaches infinity. 

To protect the privacy of the user's requested message, the $K$ strategies (corresponding to the $K$ possible messages) must be identically distributed from the perspective of each database, i.e., we must satisfy the following constraint $\forall n \in [n], \forall k \in [K]$
\begin{align}\label{privacy}
(Q_{n}^{[1]}, A_{n}^{[1]}, W_1, \ldots, W_K, Z)\sim (Q_{n}^{[k]}, A_{n}^{[k]}, W_1, \ldots, W_K, Z)
\end{align}
We say that a pair $(D,L)$ is achievable if there exists a cache enabled PIR scheme with a cache encoding, querying, and decoding functions,which satisfies the correctness and privacy constraints. The performance of a PIR scheme is characterized by the number of bits of desired information ($L$) per downloaded bit. In particular, if $D$ is the total number of downloaded bits, and $L$ is the size of the desired message, then the normalized downloaded cost is $D/L$. In other words, the PIR rate is $L/D$. 
The goal of this work is to characterize the optimal normalized download cost as a function of the storage $S$, and is defined as 
\begin{align}
D^{*}(S)= \text{min }\{D/L: (D,L) \text{ is achievable}\}.
\end{align}
The optimal PIR rate (or the capacity) is the inverse of the normalized download cost 
\begin{align}
C^{*}(S)= \text{max }\{L/D: (D,L) \text{ is achievable}\}.
\end{align}
\section{Main Results and Insights}
We first state the following Lemma which shows that the optimal download cost $D^{*}(S)$ (or the inverse of capacity $1/C^{*}(S)$) is a convex function of the storage parameter $S$. 

\begin{Lem}\label{Lemma1}
The optimal download cost $D^{*}(S)$ is a convex function of $S$. In other words, for any $(S_1, S_2)$, and $\alpha \in [0,1]$, the optimal download cost satisfies
\begin{align}
D^{*}(\alpha S_1 + (1-\alpha)S_2) \leq \alpha D^{*}(S_1) + (1-\alpha)D^{*}(S_2).
\end{align}
\end{Lem}

\textit{Proof of Lemma \ref{Lemma1}--} Let us consider two storage parameters $S_1$, and $S_2$, with optimal download costs $D^{*}(S_1)$, and $D^{*}(S_2)$ respectively using two PIR schemes, say Scheme $1$ and Scheme $2$. Let us now consider a new storage point $S= \alpha S_1 + (1-\alpha) S_2$, for which we can construct a PIR scheme as follows: we take each message $W_i$ and divide it into two independent parts $W_i = \left(W_i^{(1)}, W_i^{(2)}\right)$, where $W_{i}^{(1)}$ is of size $\alpha L$ bits, and $W_{i}^{(2)}$ is of size $(1-\alpha)L$ bits. We apply PIR Scheme $1$ on the first part of all the messages $(W_1^{(1)}, \ldots, W_K^{(1)})$, and apply PIR Scheme $2$ on the second part of all the messages $(W_1^{(2)}, \ldots, W_K^{(2)})$. The cache storage necessary for PIR scheme $1$ is $\alpha S_1L$ bits, and the download cost for this part of the message is $\alpha  D^{*}(S_1)L$ bits. Similarly,  the cache storage necessary for PIR scheme $2$ is $(1-\alpha)S_2L$ bits, and the download cost for this part of the message is $(1-\alpha)D^{*}(S_2)L$ bits. Therefore, by memory sharing between these two schemes (i.e., splitting the messages and storage and implementing two PIR schemes on independent parts of the messages), the total used cache storage is $(\alpha S_1 + (1-\alpha)S_2)L$ bits, and the total amount downloaded is $(\alpha D^{*}(S_1) + (1-\alpha)D^{*}(S_2))L$ bits. Since $D^{*}(\alpha S_1 + (1-\alpha)S_2)$ by definition is optimal download cost for storage $(\alpha S_1 + (1-\alpha)S_2)$, it must be upper bounded by the download cost of the memory sharing scheme. Hence, the following inequality follows
\begin{align}
D^{*}(\alpha S_1 + (1-\alpha)S_2) \leq \alpha D^{*}(S_1) + (1-\alpha)D^{*}(S_2),
\end{align}
which proves the convexity of $D^{*}(S)$.

\begin{Cor} \label{Corollary1}
The optimal download cost $D^{*}(S)$ is lower bounded as
\begin{align}
D^{*}(S)\leq \left(1- \frac{S}{K}\right)\left(1+ \frac{1}{N}+ \ldots + \frac{1}{N^{K-1}}\right).
\end{align}
\end{Cor}

Corollary \ref{Corollary1} follows directly from Lemma \ref{Lemma1} by noting that when $S=0$, then the optimal download cost is $D^{*}(0)= (1+ 1/N + \ldots + 1/N^{K-1})$ \cite{SunAndJaffar1}, and for $S=K$, the optimal download cost is $D^{*}(K)= 0$. Any value $S\in [0, K]$ can be written as $S = \alpha \times 0 + (1-\alpha) \times K$, i.e., $\alpha = (1-S/K)$. Hence, by Lemma \ref{Lemma1},  we have $D^{*}(S)\leq (1-S/K)D^{*}(0) + (S/K)D^{*}(K)$ which proves Corollary \ref{Corollary1}.
\\

\noindent We next state Theorem \ref{Theorem1} which is the main result of this paper. 
\begin{Theo}\label{Theorem1}
The optimal  normalized download cost $D^{*}$ of cache enabled PIR, with $K$ messages (each of size $L$ bits), $N$ databases, and a user with a normalized cache storage of $S$ is given by
\begin{align}
D^{*}(S)= \left(1- \frac{S}{K}\right)\left(1+ \frac{1}{N}+ \ldots + \frac{1}{N^{K-1}}\right).
\end{align}
\end{Theo}
\textit{Proof of Theorem \ref{Theorem1}--} In order to prove Theorem \ref{Theorem1}, we need to prove the following lower bound
\begin{align}
D^{*}(S)\geq \left(1- \frac{S}{K}\right)\left(1+ \frac{1}{N}+ \ldots + \frac{1}{N^{K-1}}\right).
\end{align}
To this end, we define $Q^{[1]}_{1:N}=\{Q_n^{[1]}, n \in [N]\}$, as the set of all $N$ queries for message $1$, $A^{[1]}_{1:N}=\{A_n^{[1]}, n \in [N]\}$, as the set of all $N$ answers for message $1$, and start with the following sequence of inequalities:
\begin{align}
&I(W_{2:K}; Q^{[1]}_{1:N}, A^{[1]}_{1:N}|Z,W_1)\nonumber\\
&= I(W_{2:K}; Q^{[1]}_{1:N}, A^{[1]}_{1:N}, W_1|Z) - I(W_{2:K}; W_1|Z)\nonumber\\
&= I(W_{2:K}; Q^{[1]}_{1:N}, A^{[1]}_{1:N}| Z) + I(W_{2:K};W_1|Q^{[1]}_{1:N}, A^{[1]}_{1:N}, Z) - I(W_{2:K}; W_1|Z)\nonumber\\
&\overset{(a)}{\leq} I(W_{2:K}; Q^{[1]}_{1:N}, A^{[1]}_{1:N}| Z) - I(W_{2:K}; W_1|Z) + o(L)\nonumber\\
&= I(W_{2:K}; A^{[1]}_{1:N}|Z, Q^{[1]}_{1:N}) - I(W_{2:K}; W_1|Z) + o(L)\nonumber\\
&=  I(W_{2:K}; A^{[1]}_{1:N}| Z, Q^{[1]}_{1:N}) - I(W_{2:K}; W_1|Z) + o(L)\nonumber\\
&=  H(A^{[1]}_{1:N}| Z, Q^{[1]}_{1:N}) - H(A^{[1]}_{1:N}| Z, Q^{[1]}_{1:N}, W_{2:K})+ o(L)\nonumber\\
&\leq   D - H(A^{[1]}_{1:N}| Z, Q^{[1]}_{1:N}, W_{2:K}) - I(W_{2:K}; W_1|Z) + o(L)\nonumber\\
&\leq  D - H(W_1, A^{[1]}_{1:N}| Z, Q^{[1]}_{1:N}, W_{2:K}) - I(W_{2:K}; W_1|Z) + o(L)\nonumber\\
&=   D - H(W_1 | Z, Q^{[1]}_{1:N}, W_{2:K}) - I(W_{2:K}; W_1|Z) + o(L)\nonumber\\
&\overset{(b)}{=}   D - H(W_1 | Z, W_{2:K}) - I(W_{2:K}; W_1|Z) + o(L)\nonumber\\
&=   D - H(W_1 | Z) + o(L),\label{eq1}
\end{align}
where (a) follows from the decodability constraint for the requested message $W_1$ from the answers, queries, and the local cache, i.e., (\ref{decodability}), and (b) follows from (\ref{privacy}). 

Hence, from (\ref{eq1}), we have
\begin{align}
D+ o(L)\geq H(W_1|Z) + I(W_{2:K}; Q^{[1]}_{1:N}, A^{[1]}_{1:N}|Z,W_1). \label{eq2}
\end{align}
In order to bound the second term in (\ref{eq2}), we next state the following Lemma, whose proof follows in a similar fashion as that in the work of Sun and Jafar \cite{SunAndJaffar1}. 
\begin{Lem}\label{Lem2}
	\begin{align}
I(W_{k:K}; Q^{[k-1]}_{1:N}, A^{[k-1]}_{1:N}|Z, W_{1:k-1}) \geq \frac{H(W_k | Z, W_{1:k-1})}{N} + \frac{I(W_{k+1;K}; Q^{[k]}_{1:N}, A^{[k]}_{1:N}| Z, W_{1:k})}{N} - o(L).\nonumber
	\end{align}
\end{Lem}
The proof of Lemma \ref{Lem2} is given in the Appendix. Using Lemma \ref{Lem2} repeatedly on the second term in (\ref{eq2}), we have the following bound on $D$ 
\begin{align}
D+o(L)&\geq H(W_{1}|Z) + \frac{H(W_2|Z, W_{1})}{N} + \frac{H(W_3|Z, W_{1:2})}{N^{2}} + \cdots + \frac{H(W_{K}|Z, W_{1:K-1})}{N^{K-1}} \nonumber\\
&= \sum_{k=1}^{K}\frac{H(W_{k} | Z, W_1, \ldots, W_{k-1})}{N^{k-1}} \nonumber\\
&= \sum_{k=1}^{K}\frac{H(W_1, \ldots, W_{k}|Z) - H(W_1, \ldots, W_{k-1}|Z)}{N^{k-1}}. \nonumber
\end{align}
We next observe that the above bound followed a sequence of steps i.e., starting with querying message $1$, followed by $2$, and up until message $K$. By following a different order (in particular, a different sequence of applications of Lemma \ref{Lem2} for another ordering of messages), we can obtain the following bound for any permutation $(\pi_1, \pi_2, \ldots, \pi_K)$ of the message set $(1, 2, \ldots, K)$. 

\begin{align}\label{perm}
D+ o(L)&\geq \sum_{k=1}^{K}\frac{H(W_{\pi_1}, \ldots, W_{\pi_k}|Z) - H(W_{\pi_1}, \ldots, W_{\pi_{k-1}}|Z)}{N^{k-1}}.
\end{align}
As an example, consider the case of $K=3$ messages and $N$ databases. For this case, we have a total of $3!=6$ bounds on $D$, given as follows:
\begin{align}
&D+ o(L)\geq H(W_1|Z) + \frac{H(W_1, W_2|Z)- H(W_1|Z)}{N} + \frac{H(W_1, W_2, W_3|Z)- H(W_1, W_2|Z)}{N^2}\nonumber\\
&D+ o(L)\geq H(W_1|Z) + \frac{H(W_1, W_3|Z)- H(W_1|Z)}{N} + \frac{H(W_1, W_2, W_3|Z)- H(W_1, W_3|Z)}{N^2}\nonumber\\
&D+ o(L)\geq H(W_2|Z) + \frac{H(W_2, W_1|Z)- H(W_2|Z)}{N} + \frac{H(W_1, W_2, W_3|Z)- H(W_2, W_1|Z)}{N^2}\nonumber\\
&D+ o(L)\geq H(W_2|Z) + \frac{H(W_2, W_3|Z)- H(W_2|Z)}{N} + \frac{H(W_1, W_2, W_3|Z)- H(W_2, W_3|Z)}{N^2}\nonumber\\
&D+ o(L)\geq H(W_3|Z) + \frac{H(W_3, W_1|Z)- H(W_3|Z)}{N} + \frac{H(W_1, W_2, W_3|Z)- H(W_3, W_1|Z)}{N^2}\nonumber\\
&D+ o(L)\geq H(W_3|Z) + \frac{H(W_3, W_2|Z)- H(W_3|Z)}{N} + \frac{H(W_1, W_2, W_3|Z)- H(W_3, W_2|Z)}{N^2}.\nonumber
\end{align}
Summing up these $3!$ bounds, and dividing by $3!$, we obtain
\begin{align}
&D+ o(L) \nonumber\\
&\geq \frac{1}{N^{0}}\left(\frac{(H(W_1|Z) + H(W_2|Z) + H(W_3|Z))}{{{{3}\choose{1}}}}\right) \nonumber\\
&\quad + \frac{1}{N^{1}}\Bigg( \frac{(H(W_1,W_2|Z) + H(W_2, W_3|Z) + H(W_1, W_3|Z))}{{{3}\choose{2}}} - \frac{(H(W_1|Z) + H(W_2|Z) + H(W_3|Z)}{{{3}\choose{1}}} \Bigg) \nonumber\\
&\quad + \frac{1}{N^{2}}\Bigg( \frac{H(W_1,W_2,W_3|Z)}{{{3}\choose{3}}} - \frac{(H(W_1,W_2|Z) + H(W_2,W_3|Z) + H(W_1, W_3|Z)}{{{3}\choose{2}}} \Bigg). \nonumber
\end{align}
For the general case of $K$ messages, we take all possible $K!$ permutations of (\ref{perm}), sum up all the $K!$ bounds, and divide by $K!$ to obtain the following lower bound on $D$:
\begin{align}
D+ o(L)&\geq \sum_{k=1}^{K}\frac{1}{N^{k-1}}\left(
\frac{\sum_{\mathcal{S}\subseteq [1:K]: |\mathcal{S}|=k}H(W_{\mathcal{S}}|Z)}{{{K}\choose{k}}} -
\frac{\sum_{\mathcal{S}\subseteq [1:K]: |\mathcal{S}|=(k-1)}H(W_{\mathcal{S}}|Z)}{{{K}\choose{k-1}}} 
\right).\label{eq4}
\end{align}
Before proceeding further, we next state the \textit{conditional version} of Han's inequality on the entropy of subsets of random variables which plays a key role in obtaining the tight lower bound on $D$. 

\textbf{Han's Inequality--} Consider a set of $K$ random variables $(W_1, W_2, \ldots, W_K)$, and let $W_\mathcal{S}$ denote the subset $W_{\mathcal{S}}= \{X_i: i \in \mathcal{S}\}$. Let 
\begin{align}
\mu_{k}\triangleq \frac{1}{{{K}\choose{k}}}\sum_{\mathcal{S} \subseteq [1:K]: |\mathcal{S}|=k}\frac{H(W_{\mathcal{S}}|Z)}{k}, \label{mudef}
\end{align}
then Han's inequality \cite{CoverThomas} (Chapter $17$, Theorem $17.6.1$). states the following:
\begin{align}
\mu_{1}\geq \mu_{2} \geq \cdots \geq \mu_{K}.\label{Han}
\end{align}

 Using this notation (\ref{mudef}) of $\mu_{k}$ in (\ref{eq4}), we proceed to lower  bound $D$ as follows:
\begin{align}
D + o(L)&\geq \sum_{k=1}^{K}\frac{1}{N^{k-1}}\left(
\frac{\sum_{\mathcal{S}\in [K]: |\mathcal{S}|=k}H(W_{\mathcal{S}}|Z)}{{{K}\choose{k}}} -
\frac{\sum_{\mathcal{S}\in [K]: |\mathcal{S}|=(k-1)}H(W_{\mathcal{S}}|Z)}{{{K}\choose{k-1}}} 
\right)\nonumber\\
&\overset{(a)}{=}  \sum_{k=1}^{K}\frac{1}{N^{k-1}}\left(
k\mu_{k} - (k-1)\mu_{k-1}
\right)\nonumber\\
&= \mu_1 + \frac{1}{N}(2\mu_2 - \mu_1) + \frac{1}{N^{2}}(3\mu_3 - 2\mu_2) + \frac{1}{N^{3}}(4\mu_4 - 3\mu_3) + \ldots + \frac{1}{N^{K-1}}(K\mu_K) \nonumber\\
&\overset{(b)}{=} \mu_1 \left(1-\frac{1}{N}\right) + (2\mu_2) \left(\frac{1}{N} - \frac{1}{N^{2}}\right) + (3\mu_3) \left(\frac{1}{N^{2}} - \frac{1}{N^{3}}\right) \ldots + K\mu_K \left(\frac{1}{N^{K-1}}\right)\nonumber\\
&\overset{(c)}{\geq} \mu_K \left(1-\frac{1}{N}\right) + (2\mu_K) \left(\frac{1}{N} - \frac{1}{N^{2}}\right) + (3\mu_K) \left(\frac{1}{N^{2}} - \frac{1}{N^{3}}\right) \ldots + K\mu_K \left(\frac{1}{N^{K-1}}\right)\nonumber\\
&= \mu_K \left(1+ \frac{1}{N}+ \frac{1}{N^{2}}+ \ldots + \frac{1}{N^{K-1}}\right)\nonumber
\end{align}
\begin{align}
&\overset{(d)}{=} \frac{H(W_1, W_2, \ldots, W_K |Z)}{K}\left(1+ \frac{1}{N}+ \frac{1}{N^{2}}+ \ldots + \frac{1}{N^{K-1}}\right)\nonumber\\
&= \frac{H(W_1, W_2, \ldots, W_K,Z) - H(Z)}{K}\left(1+ \frac{1}{N}+ \frac{1}{N^{2}}+ \ldots + \frac{1}{N^{K-1}}\right)\nonumber\\
&\overset{(e)}{=} \frac{H(W_1, W_2, \ldots, W_K) - H(Z)}{K}\left(1+ \frac{1}{N}+ \frac{1}{N^{2}}+ \ldots + \frac{1}{N^{K-1}}\right)\nonumber\\
&\overset{(f)}{=} \frac{(KL - SL)}{K}\left(1+ \frac{1}{N}+ \frac{1}{N^{2}}+ \ldots + \frac{1}{N^{K-1}}\right)\nonumber\\
&= L\left(1-\frac{S}{K}\right)\left(1+ \frac{1}{N}+ \frac{1}{N^{2}}+ \ldots + \frac{1}{N^{K-1}}\right),\label{eq6}
\end{align}
where $(a)$ follows by the definition of $\mu_k$ in (\ref{mudef}), $(b)$ follows by re-arranging and collecting the coefficients of $\mu_k$'s for $k=1, 2, \ldots, K$. The step $(c)$ is the most critical one and that follows from Han's inequality (\ref{Han}), which states that $\mu_1\geq \mu_2 \ldots \geq \mu_K$, and we lower bound each one of the $\mu_k$'s by $\mu_K$. Step $(d)$ follows from the fact that $\mu_K = H(W_1, W_2, \ldots, W_K | Z)/K$, and step $(e)$ is due to the fact that the cache content $Z$ is a function of the messages, i.e., $H(Z|W_1, W_2, \ldots, W_K)=0$. Finally, $(f)$ follows from the cache storage constraint, i.e., $H(Z)=SL$, and the fact that the messages are independent and each message has entropy $L$. 
Hence, from (\ref{eq6}), we have the proof for the lower bound 
\begin{align}
\frac{D}{L}\geq \left(1-\frac{S}{K}\right)\sum_{k=1}^{K}\frac{1}{ N^{k-1}} -\frac{o(L)}{L}.
\end{align}
Taking the limit $L\rightarrow \infty$, we have the proof of Theorem \ref{Theorem1}.
\vspace{-15pt}
\section{Conclusions and Discussion}
\vspace{-5pt}
In this work, we characterized the capacity of cache enabled private information retrieval  (PIR) and it was shown to be the inverse of $(1- \frac{S}{K})\left(1+ \frac{1}{N}+ \ldots + \frac{1}{N^{K-1}}\right)$, as a function of $(N, K)$ and the storage parameter $S$. The key technical contribution of this work is the  converse, i.e., a lower bound on the download cost as a function of storage $S$ which shows that a simple memory sharing scheme is information-theoretically optimal. The new ingredient in the proof (in addition to those in \cite{SunAndJaffar1}) include an interesting use of Han's inequality to bound the conditional entropies of subsets of messages conditioned on the cache content $Z$. We would like to remark that for the cache aided PIR problem considered in this work, it was assumed that the cached content $Z$ is publically known, i.e., it is revealed to all the databases. 

As a future direction, an interesting variant of the problem could be \textit{dynamic} cache aided PIR, in which the cached content of the user is not revealed to the databases. By leveraging the unknown cache, it is entirely possible to improve the capacity beyond memory sharing. However, in such a scenario, once the \textit{unknown} cache is used, the user would need to update/refresh its cached contents (either through the databases themselves, or through some other trusted mechanism which keeps the cached content \textit{essentially random} from the perspective of each database). Thus, the total download cost would comprise of two elements: the cost of retrieving the desired message, and the cost of refreshing the cached contents for efficient PIR in the future. It would be interesting to see if one can improve upon memory sharing even after accounting for the total cost of download and cache refreshment, and  characterize this general tradeoff. 

\section{Appendix: Proof of Lemma \ref{Lem2}}
To prove Lemma \ref{Lem2}, we start with the following sequence of inequalities
\begin{align}
N I(W_{k:K}; Q^{[k-1]}_{1:N}, A^{[k-1]}_{1:N}|Z, W_{1:k-1})&\geq \sum_{n=1}^{N} I(W_{k:K}; Q^{[k-1]}_{n}, A^{[k-1]}_{n}| Z, W_{1:k-1})\nonumber\\
& \overset{(\ref{privacy})}{=} \sum_{n=1}^{N} I(W_{k:K}; Q^{[k]}_{n}, A^{[k]}_{n}|Z, W_{1:k-1})\nonumber\\
& \geq \sum_{n=1}^{N} I(W_{k:K}; A^{[k]}_{n}| Z, W_{1:k-1}, Q^{[k]}_{n})\nonumber\\
& \overset{(\ref{answer})}{=} \sum_{n=1}^{N} H(A^{[k]}_{n}| Z, W_{1:k-1}, Q^{[k]}_{n})\nonumber\\
& \geq \sum_{n=1}^{N} H(A^{[k]}_{n} | Z, W_{1:k-1}, Q^{[k]}_{1:N}, A^{[k]}_{1:n-1})\nonumber\\
& = I( W_{k;K}; A^{[k]}_{1:N}|Z, W_{1:k-1}, Q^{[k]}_{1:N})\nonumber\\
& \overset{(\ref{privacy})}{=} I( W_{k;K}; Q^{[k]}_{1:N}, A^{[k]}_{1:N}|Z, W_{1:k-1})\nonumber\\
&  \overset{(\ref{decodability})}{=} I( W_{k;K}; W_k, Q^{[k]}_{1:N}, A^{[k]}_{1:N}| Z, W_{1:k-1}) - o(L)\nonumber\\
&= I(W_{k;K}; W_k|Z, W_{1:k-1}) + I(W_{k;K}; Q^{[k]}_{1:N}, A^{[k]}_{1:N}| Z, W_{1:k})- o(L)\nonumber\\
&= H(W_k | Z, W_{1:k-1}) + I(W_{k+1;K}; Q^{[k]}_{1:N}, A^{[k]}_{1:N}| Z, W_{1:k}) - o(L).\nonumber
\end{align}
Hence, we have 
\begin{align}
I(W_{k:K}; Q^{[k-1]}_{1:N}, A^{[k-1]}_{1:N}|Z, W_{1:k-1}) \geq \frac{H(W_k | Z, W_{1:k-1})}{N} + \frac{I(W_{k+1;K}; Q^{[k]}_{1:N}, A^{[k]}_{1:N}| Z, W_{1:k})}{N} - o(L),\nonumber
\end{align}
which completes the proof of Lemma \ref{Lem2}. 

\bibliographystyle{unsrt}
\bibliography{paper}
\end{document}